\documentclass[11pt]{article}
\usepackage{times}
\usepackage{geometry}
\usepackage[utf8]{inputenc}
\usepackage{enumitem,amssymb}
\usepackage{graphicx}
\usepackage{fancyhdr}
\usepackage{aas_macros}
\usepackage{mdframed} 
\usepackage[authoryear]{natbib}
\usepackage{hyperref}

\setcitestyle{authoryear,open={(},close={)}}

\mdfdefinestyle{theoremstyle}{
innertopmargin=\topskip,}
\mdtheorem[style=theoremstyle]{lrptextbox}{}

\begin{document}


%
%

\thispagestyle{empty}

\begin{flushleft}
{$\phantom{AAAAAAAA}$ 
\bf \LARGE The cosmic origin and evolution of the elements}\\
{$\phantom{AAAAAA}$ White paper for the Canadian Long Range Plan for Astronomy and Astrophysics 2020}
\end{flushleft}

\noindent
R.~Fern\'andez$^{1}$, F.~Herwig$^2$, S.~Safi-Harb$^3$,
I.~Dillmann$^{2,4}$, K.~A.~Venn$^2$, 
B.~C\^ot\'e$^5$, C.~O.~Heinke$^1$, 
E.~Rosolowsky$^1$, T.~E.~Woods$^{6}$, 
D.~Haggard$^7$, L.~Lehner$^{8,9}$, 
J.~J.~Ruan$^7$, D.~M.~Siegel$^{7,10}$, 
J.~Bovy$^{11}$, A.~A.~Chen$^{12}$, A.~Cumming$^7$,
B.~Davids$^{4,13}$, M.~R.~Drout$^{11}$, R.~Kr\"ucken$^{4,14}$

{\smallskip}
\noindent
{\small 
(1)~Univ.~of~Alberta 
(2)~Univ.~of~Victoria 
(3)~Univ.~of~Manitoba
(4)~TRIUMF 
(5)~Konkloy~Observatory 
(6)~NRC~Herzberg 
(7)~McGill~Univ. 
(9)~Perimeter~Institute 
(9)~Univ.~of~Waterloo 
(10)~Univ.~of~Guelph
(11)~Univ.~of~Toronto
(12)~McMaster~Univ. 
(13)~Simon~Fraser Univ. 
(14)~Univ. of British Columbia
}

\section*{Executive Summary}

The origin of many elements of the periodic table remains an unsolved problem.
Many nucleosynthetic channels are broadly understood, mostly those involving
the quiescent stellar evolution phases that can be reliably modeled in
spherical symmetry. However, significant uncertainties remain regarding our
understanding of certain groups of elements, such as the intermediate and rapid
neutron-capture processes, the p-process, or the origin of odd-Z elements in
the most metal-poor stars.

Beyond completing our understanding of the origin of the elements, nuclear
astrophysics aims to provide reliable predictions for when, where, and which
elements are released from dying stars. Coupled with galactic and extragalactic
observations of stellar abundances, this approach, known as galactic
archaeology, can provide key insights into how galaxies form and evolve.
However, the predictive power and fidelity of our simulations are in many cases
insufficient to fully deliver on this vision, partly due to the intrinsically
three-dimensional (3D) nature of the stellar processes involved. Only recently
can dynamic events in the evolution of massive stars be tracked with
high-fidelity, 3D hydrodynamic simulations; these have already led to scenarios
that can explain the origin of odd-Z elements. Also, asteroseismology is
emerging as a powerful tool to validate these new stellar models which will
ultimately create 3D progenitors for 3D supernova explosion simulations that
are key to understanding the diverse nature of supernova remnant and stellar
abundance observations.

On the observational side, the detection in 2017 of a neutron star merger in
gravitational waves started the era of multi-messenger gravitational wave
astronomy, with multi-wavelength follow-up of that event resulting in the
confirmation of neutron star mergers as sites for the rapid neutron capture
process, given the broad agreement with theoretical predictions. Since then,
gravitational wave observatories have improved in sensitivity, yielding a
higher rate of detections. This trend will continue to increase as future
upgrades and new facilities come online. Larger sample sizes will test our
understanding of these events, requiring reliable predictions, which are
computationally very challenging.

Emerging and future multi-wavelength surveys are delivering large data sets of
stellar abundances. These have the potential to provide a new window into
galactic formation and evolution processes, if they can be paired with reliable
stellar yield models. A key element to fulfill this vision is to address the
substantial uncertainties in modeling the stellar atmospheres required to
determine abundances reliably. Here, 3D hydrodynamic and non-LTE effects still
again provide major challenges.

Unraveling the origin of the elements in the context of galaxy formation and
evolution requires, on the Astrophysics side, the interplay between
observations of stellar abundances in large surveys, detailed studies of
individual objects such as supernova remnants, multi-messenger transient
follow-up, new generation of 3D hydrodynamic stellar evolution models, and
simulations of explosive events using neutrino radiation-magnetohydrodynamics
in numerical relativity with nuclear processes. A key ingredient to
interpreting observations and generating theoretical predictions is a wide
array of fundamental nuclear data properties, such as nuclear reaction cross
sections and the equation of state at high densities. For much of the nuclear
data of unstable species required, however, only theoretical predictions are
available. Experiments with unstable (rare) isotope beams are now becoming
possible in the regime relevant to address this critical nuclear data need.
{\bf Nuclear astrophysics is therefore an interdisciplinary research frontier that
integrates all of these sub-fields of Astrophysics and Experimental Physics.}

Sustaining Canadian leadership on the observational side in the next decade
will require access to transient and non-transient surveys like LSST, SKA, or
MSE, support for target-of-opportunity observing in current and future Canadian
telescopes, and participation in next-generation X-ray telescopes such as
ATHENA. On the theory side, state-of-the-art predictions for the next decade
will require an ambitious succession to the Niagara supercomputer to support
large parallel jobs.

The lack of funding for postdoctoral researchers and of funding envelopes for
such an interdisciplinary collaboration prevents Canadian scientists from
competing on a level-playing field with international groups, as existing
funding programs do not meet the needs of the field. We propose a funding
instrument for postdoctoral training that reflects the interdisciplinary nature
of nuclear astrophysics research. We also propose the creation of a national
collaborative funding program that allows for joint projects and workshop
organization, increasing ties between these communities.

Canada has a long tradition of leadership in nuclear astrophysics, dating back
to the work of Alastair Cameron at Chalk River Laboratory in the 1950s. Work by
the Canadian community up to 2010 has been comprehensively summarized in the
2010 white paper \emph{Nuclear Astrophysics in Canada}\footnote{Available at
\url{https://astro.triumf.ca/canadian-nuclear-astrophysics-white-paper}}. Since
then, faculty hires in Astronomy and Physics Departments have further boosted
activity in the field, including transient observations (Drout, Gaensler,
Haggard, Hlo\v zek, Sivakoff) and theory (East, Fern\'andez, Siegel), as well
as survey science on galactic nucleosynthesis (Bovy, Venn), and nuclear
experiments (Christian). 

This white paper provides a brief overview of recent activity in the community,
highlighting strengths in each specific sub-field. We then provide
recommendations to improve interdisciplinary collaboration. Other white papers
that relate to this topic are led by Ruan (E035, transients), Spekkens (E042,
SKA), Hlo\v zek (E023, LSST), Venn (E015, Machine Learning), C\^ot\'e (E045,
Gemini), and Hoffman (E048, Colibr\`i).

\section{The formation and evolution of the elements in stars and galaxies}
\label{s:science1}

\subsection{Observational diagnostics}

\begin{figure*}[t]
\includegraphics*[width=0.42\textwidth]{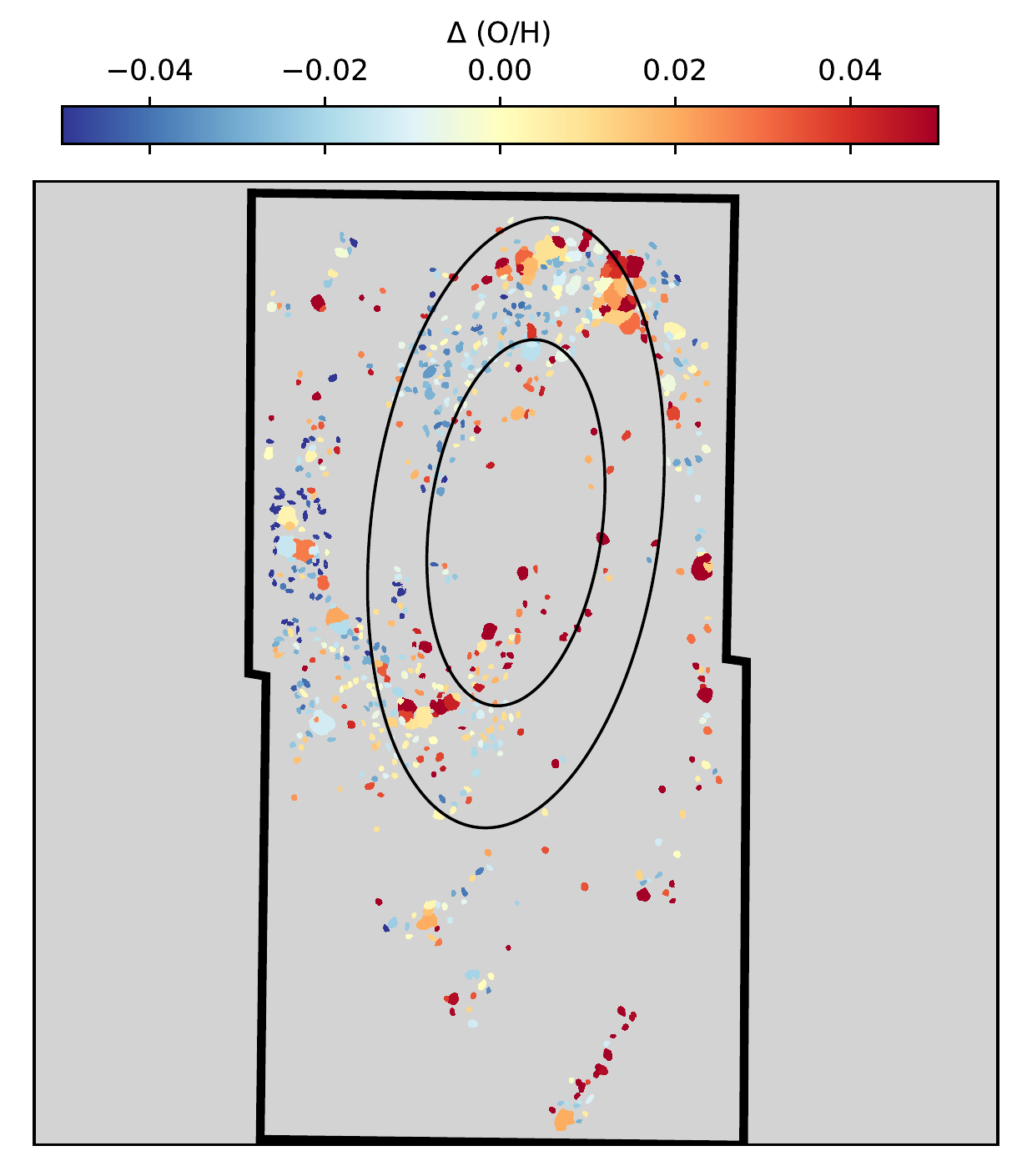}
\includegraphics*[width=0.55\textwidth]{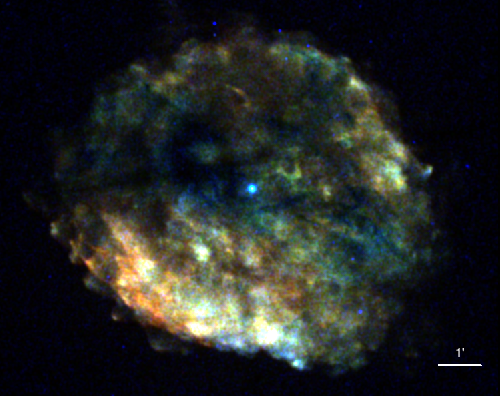}
\caption{\emph{Left:} Abundance map of NGC 3627 using VLT/MUSE from the PHANGS
project \citep{kreckel_2019}.  The image shows the abundances of HII
regions relative to the azimuthally averaged abundance gradient, indicating
enhanced oxygen abundances in regions of vigorous star formation at the bar
ends of the galaxy.  Studies of metallicity fluctuations constrain the
enrichment process and how newly created metals are integrated into future
generations of stars.  \emph{Right}: RGB \textit{Chandra} X-ray image of the
supernova remnant RCW~103 hosting a central compact object that once behaved
like a magnetar (\citealt{Braun2019}).  Imaging and spectral analysis of the
chemical elements in supernova remnants provides clues to the supernova
explosion mechanism and its engine.}
\label{f:obs}
\end{figure*}

The metallicity of the ISM is a direct tracer of galactic enrichment, measuring
the abundances in the material that is being incorporated into the
currently-forming stellar population.  Typically, these abundances are measured
from optical emission line studies of HII regions, but careful analysis of
dust-to-gas ratios and/or molecular line emission also provide some insights
into enrichments \citep{wilson94}.  The original observations of abundance
gradients came from extragalactic HII regions where abundance gradients in
galaxies have been known for decades \citep{searle71}.  These gradients
constrain the models of galaxy growth, with the ISM gradients providing
complementary information to stellar metallicity.  Specifically, ISM studies
show the enrichment of gas that is being incorporated into the current
generation of stars. The ISM should also show local enrichment by supernovae,
and searches for these signatures are becoming possible thanks to refined
methods for determining HII region metallicity as well as the deployment of 2D
integral field unit spectroscopy from SITELLE \citep{signals}, MANGA
\citep{thorp19} and MUSE (Figure~\ref{f:obs}).  These signatures of local
enrichment provide observational constraints on abundance yield from stars,
outflow rates driven by star formation, and mixing processes within the ISM
\citep[e.g.,][]{armillota18}.

While ISM studies provide unique insights into the current production and
distribution of chemical elements within galaxies, stellar abundances derived
from spectroscopy probe the complete production history from the birth of
galaxies in the early Universe to the current time (\citealt{venn04}). The most
metal-poor stars found in the Galactic halo and in dwarf galaxies can even
carry the chemical fingerprints of the first stars that formed in the Universe
(\citealt{frebel15_review}). The Canadian efforts on the Pristine Survey
\citep{pristineIV,venn2019}, the new Gemini Observatory blue-sensitive GHOST
spectrograph \citep{GHOST2016}, and the Maunakea Spectroscopic Explorer
(\citealt{mse19}) will put Canadian scientists at the forefront of nuclear
astrophysics research on the rise of the elements in the early Universe. The
Solar System composition is another powerful diagnostic as it includes all
stable elements and isotopes (\citealt{asplund09}). It is the only system where
a large variety of nuclear processes can be probed via isotopic ratios, in turn
creating solid links between astronomy, theoretical nucleosynthesis, and
experimental nuclear physics at facilities such as TRIUMF
(\S\ref{Sec:TRIUMF-Astro}).

\subsection{Nuclear processes in galaxy evolution}
\label{s.galevol}

Galaxy formation and evolution plays a central role in studying the evolution
of the elements (\citealt{mackereth19}). The mechanical energy and radiation
released into the ISM by stars shape the internal structure of galaxies and
regulate its star formation (e.g., \citealt{hopkins18}). This energy deposition
is triggered by various nuclear processes, and because those processes are
responsible for creating the elements and isotopes, the evolution of galaxies
and the evolution of the elements are intrinsically linked together.
Understanding the origin of matter in the Universe therefore requires
state-of-the-art 3D hydrodynamic simulations of galaxies, and high-performance
computing facilities such as the Niagara supercomputer cluster at Compute
Canada (\S\ref{s:recomm}). 

Canadian researchers are leading international collaborative efforts to combine
expertise and create connections between nuclear astrophysics and galaxy
simulations (e.g., \citealt{cote18_iprocess}).  Including our nuclear
astrophysics expertise into cosmological simulations of galaxies (e.g.,
\citealt{starkenburg17}) is necessary to best interpret the chemical abundances
of metal-poor stars and to maximize the scientific returns of major Canadian
investments such as Pristine, GHOST, and MSE. In addition, this will create an
innovative bridge between the physics of the early Universe and Canadian
nuclear physics experiments.

The foundation of chemical evolution in galaxies is the life-cycle of stars and
the mixing of elements within the galactic gas (e.g., \citealt{nomoto13}).  The
next frontier is a better quantification of inhomogeneous mixing within the
turbulent and multi-phase ISM (e.g., \citealt{rennehan19}), and a better
understanding of the physical mechanisms by which chemical elements are
transferred from one generation of stars to another. Numerically addressing the
fundamental interactions between stars and their environment will complement
the unique observational surveys pursued with SITELLE at the Canada France
Hawaii Telescope.

\subsection{Multi-dimensional stellar models and nucleosynthesis}
\label{s:uvic_stellar_hydro}
\begin{figure*}
\includegraphics*[width=0.49\textwidth]{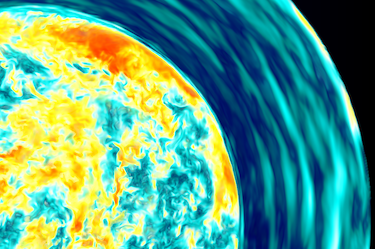}
\includegraphics*[width=0.51\textwidth]{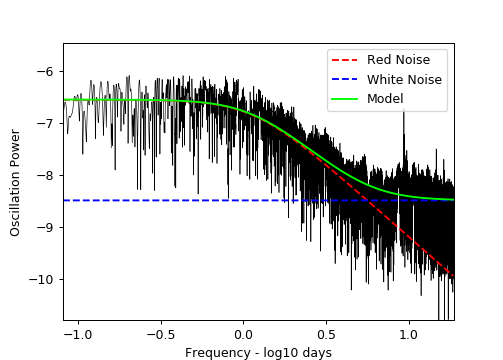}
\caption{Left: A part of a thin central-plane slice of the horizontal velocity
component of a 3D simulation of core convection in a massive star. In the
upper-right part of the partial image the internal gravity waves in the stable
layer can be discerned. Right: The associated spectrum of the internal gravity
waves in the simulation are reproducing astonishingly well the asteroseismic,
observed low-frequency excess recently discovered in massive stars by
\citet{Bowman:2019ka}.}
\label{f:stellar_abundances}
\end{figure*}

Canadian researchers are leading with their collaborators the frontier of 3D
hydrodynamic simulations of stellar convection in the late phases of low-mass
\citep{Herwig:2014cx} and massive stars \citep{Jones:2017kc}. Such simulations
provide key input into comprehensive stellar model and yield calculations
\citep{Ritter:2018kb} that are input to the galaxy models described in
\S\ref{s.galevol}. 

Realistic 3D simulations of convection interacting with nuclear production
processes activate unique new nucleosynthesis pathways away from stability,
thereby providing the astrophysical environment for the formation of rare
isotopes that are now possible to probe experimentally at TRIUMF-ISAC
(\S\ref{Sec:TRIUMF-Astro}). For example, 3D simulations have  been instrumental
in establishing the new role of the intermediate n-capture process to explain
abundance patterns observed in metal-poor stars \citep{2019MNRAS.488.4258D}.
The urgent need of new nuclear physics data has already triggered  nuclear
physics experiments at TRIUMF (\S\ref{Sec:TRIUMF-Astro}) and at other major
nuclear physics labs (e.g.\ NSCL/MSU/JINA).

A very exciting new development is the prospect of comparing the 3D simulations
outputs directly with space-based asteroseismology observations from TESS, and
in the future Plato. In 2018, as part of the early user access to the Compute
Canada Niagara super-computer, the UVic group performed the highest-resolution
3D simulation of core convection in a massive star\footnote{YouTube movies:
\url{https://bit.ly/2HzTKtw}; R\&D  Magazine: \url{https://bit.ly/2K95Gac}}.
Subsequent analysis of the frequency spectrum of the internal gravity waves
excited in the stable layer above the core convection showed astonishing
agreement with recently discovered asteroseismic features observed in massive
stars (Figure~\ref{f:stellar_abundances}). 

\newpage
\section{Explosive transients}
\label{s:science2}

\subsection{Neutron star mergers: gravitational waves and $r$-process nucleosynthesis}

\begin{figure*}
\includegraphics*[width=0.4\textwidth]{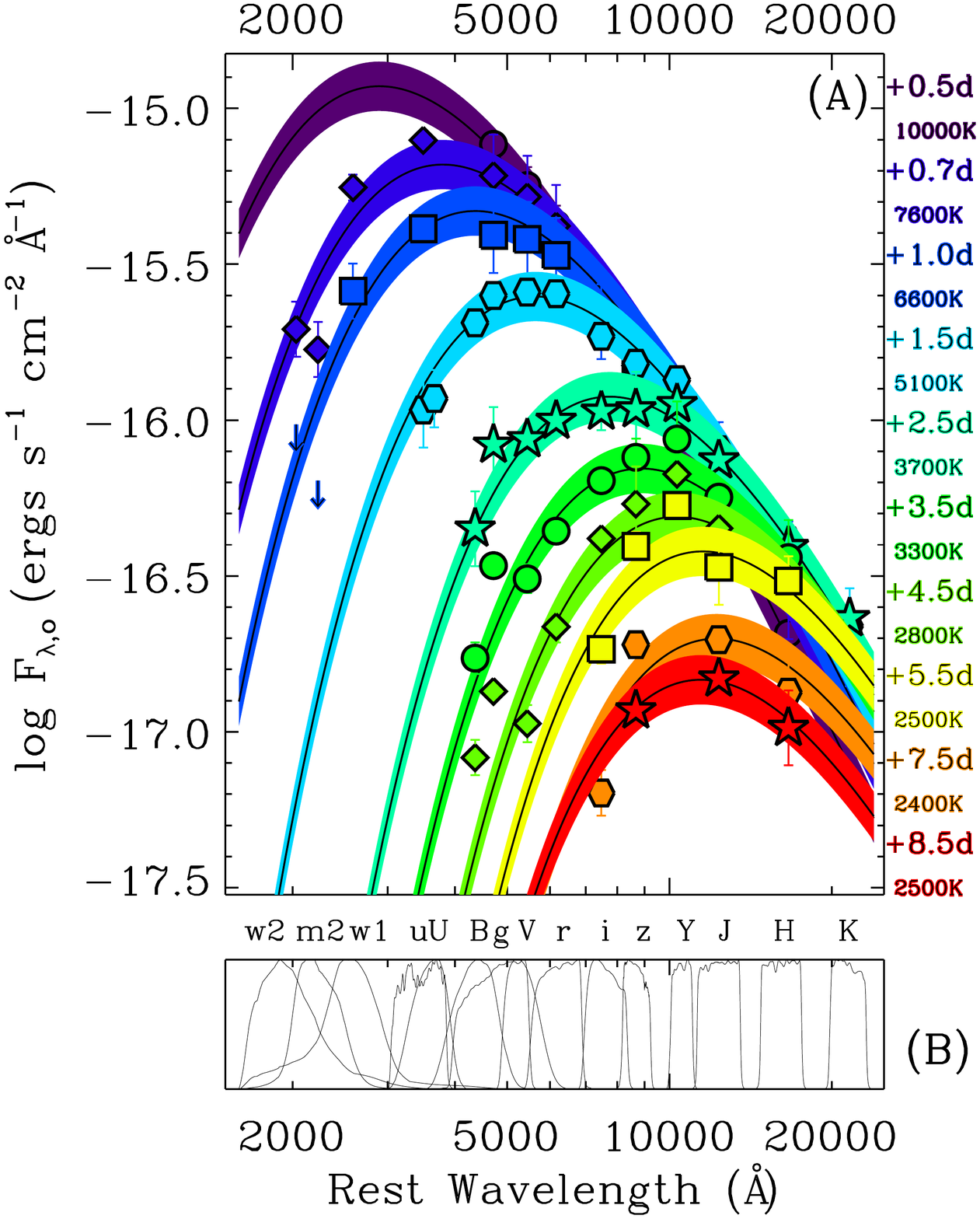}
\includegraphics*[width=0.6\textwidth]{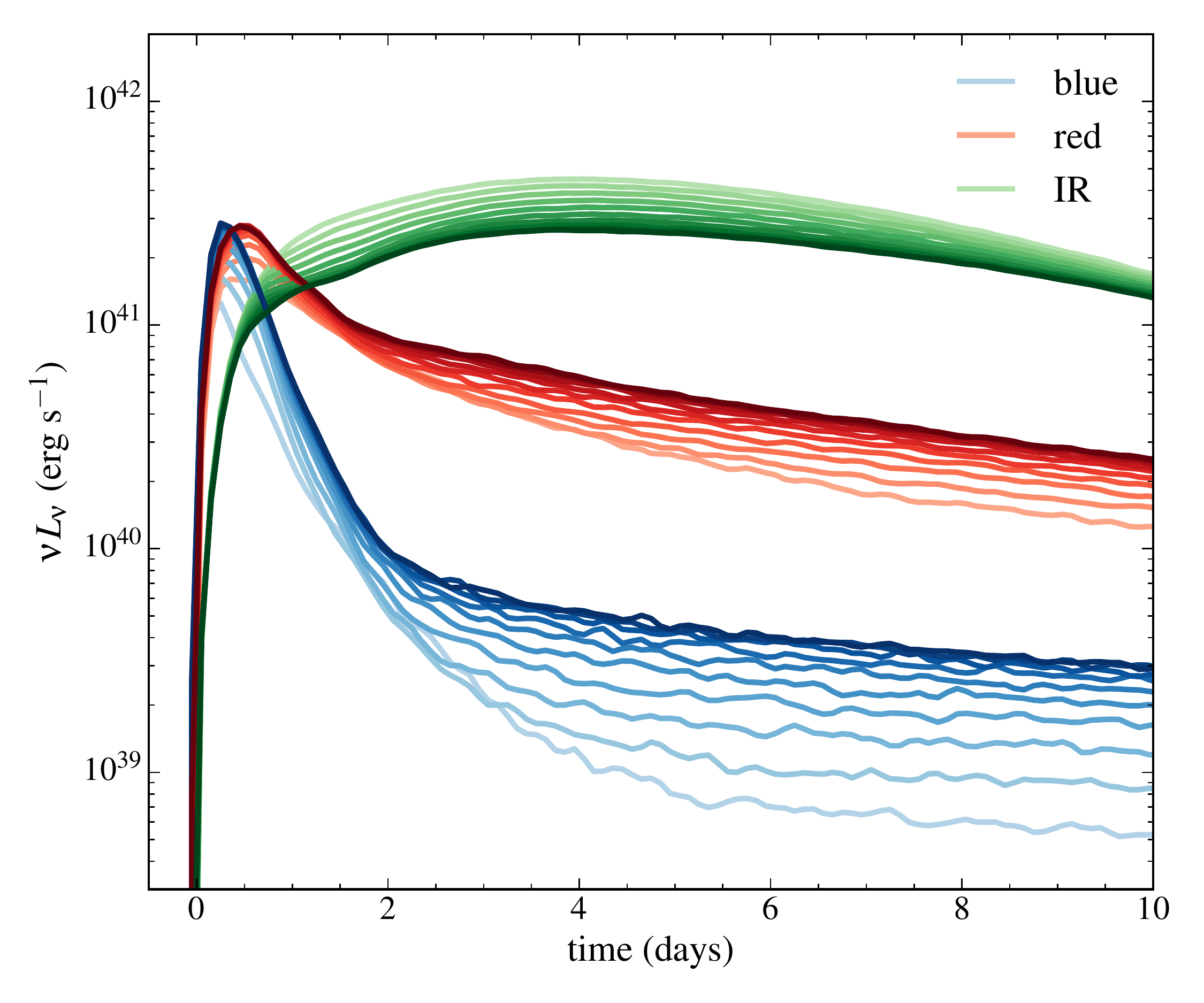}
\caption{\emph{Left}: optical/IR spectral energy distribution for the kilonova
from GW170817 as a function of time, showing the shift from blue optical to
near infrared within a few days (from \citealt{drout_2017}). \emph{Right:}
kilonova light curve predictions for generic BH-NS mergers in various spectral
bands (from \citealt{fernandez_2017}). The rapid color evolution of the light
curve over a few days is a consequence of the sensitivity of the optical
opacity to the presence of heavy $r$-process elements.}
\label{f:gw170817}
\end{figure*}

Compact object mergers are the main target for direct detection in
gravitational waves by ground based interferometers. Neutron star mergers in
particular had long been predicted to be a site for the $r$-process
\citep{lattimer_1974}, and in 2017 they were first detected in both
gravitational and electromagnetic waves (\citealt{gw170817_mm}; multiple
Canadian co-authors). The observations were in broad agreement with
state-of-the-art theoretical predictions for the kilonova counterpart,
providing evidence for the operation of the $r$-process in these events
(Figure~\ref{f:gw170817}).

At present, the LIGO and Virgo observatories are carrying out their 3rd
observing run, and will soon be joined by the Japanese KAGRA observatory.
Future observing runs and upgrades are planned for the next decade, with steady
increases in sensitivity \citep{ligo_obs_plans}.  Extensive electromagnetic
follow-up capabilities are in place worldwide, with public alerts issued within
seconds of a gravitational wave detection. 
The Canadian community played a key role in the follow-up of GW170817 (e.g.,
\citealt{drout_2017,haggard_2017}) as well as in formulating theoretical
predictions for and interpretation of electromagnetic counterparts (e.g.,
\citealt{lehner_2016,siegel_2017,fernandez_2017,cote18_ligo}). At present,
Canadian teams have standing follow-up allocations on Gemini, CFHT, Chandra and
the Jansky VLA. For more information, see white paper led by Ruan (E035).
Reliable theoretical predictions for the electromagnetic signal and
nucleosynthesis yield from neutron star mergers in the next decade will require
simulations that combine numerical relativity,
neutrino-radiation-magnetohydrodynamics, and nuclear processes. These
calculations are carried out at the limit of current capabilities, requiring
compute clusters of the scale of Niagara or even larger (\S\ref{s:recomm}).

\subsection{Type Ia Supernovae: progenitors and remnants}

Type Ia supernovae (SNe Ia) are a vital component of nucleosynthesis, in
particular being responsible for much of the iron in the Universe
\citep[e.g.,][]{Hitomi2017}.  They are now firmly understood to be the
thermonuclear explosions of carbon-oxygen white dwarfs (WD) \citep{MMN14}.
However, it remains unknown whether they arise due to one hot and luminous WD
growing slowly via accretion and surface nuclear-burning until reaching $\sim$
$M_{\rm{Ch}}$ (``single-degenerates'') or via the interaction and merger of two
WDs (``double-degenerate''), or some combination thereof. While there has been
great progress in modelling binary populations \citep[e.g.,][]{Chen14} and WD
explosions \cite[e.g.,][]{Zhu15}, fundamental uncertainties in binary stellar
evolution \citep[e.g., common envelope physics,][]{Ivanova13} and explosion
simulations have held back our understanding of how often these scenarios
occur, and whether they produce explosions which resemble SNe~Ia.

A number of recent breakthroughs in modelling WD accretion
\citep[e.g.,][]{Denissenkov17} and the impact accreting WDs have on their
environment (e.g., \citealt{Woods17}) have begun to disfavour the classic
single-degenerate model as the dominant channel. On the other hand, however,
near-$M_{\rm{Ch}}$ explosions consistent with a single-degenerate origin appear
to be essential sites for the production of $^{48}$Ca and other neutron-rich
isotopes \citep[e.g.,][]{ST17}; abundance measurements of the Perseus cluster
from the X-ray telescope {\it Hitomi} favour a mixture of sub-$M_{\rm{Ch}}$ and
near-$M_{\rm{Ch}}$ explosions \citep{Hitomi2017}. 

Going forward, multi-wavelength observations of supernova remnants (SNRs) can
provide a powerful means of assessing the viability of differing progenitor
scenarios on a case-by-case basis, while providing a direct probe of the heavy
elements synthesized in their explosions. Observations of optical emission
lines (both from space, e.g., {\it HST}, and the ground, e.g., CFHT, Gemini) in
the vicinity of SNRs can provide a powerful constraint on the temperature and
luminosity of their progenitors (e.g., \citealt{GW19}), while simultaneously
providing crucial insights into the long-term evolution of the remnant, the
physics of the shock, and the nature of the surrounding ISM
\citep[e.g.,][]{BR17}. At the same time, X-ray spectroscopy provides an
especially powerful probe of the yields of SNe Ia, and comparison with
explosion models allows one to determine the mass and structure of the WD at
the time of explosion \citep[e.g.,][]{Dave17}.  In the next decade,
multi-wavelength spectroscopy from the optical to gamma rays, high-resolution
imaging, and improved modelling will all be essential to definitively identify
the viability and relative contributions of differing progenitor models. In
particular, the availability of quantum calorimeter (e.g., XARM, Athena) X-ray
observations of SNRs will provide a qualitative improvement of our
understanding of element distribution in these explosions (\S\ref{s:recomm}). 

\subsection{Core-collapse supernovae: progenitors, explosion, and remnants}
\label{s.CCSN}

The explosions of massive stars play a key role in the evolution of the
Universe, by enriching it with heavy elements. Recent progress in understanding
the explosion mechanism has revealed that asphericities originating in
pre-supernova stellar convection constitute a major uncertainty in modelling
the supernova explosion (e.g., \citealt{janka_2016}).  3D simulations
(\S\ref{s:uvic_stellar_hydro}) can now realistically characterize the emergence
of such asphericities in dynamic mergers of convective O- and C-burning shells
\citep{Andrassy:2018wy}, and the associated dynamic nucleosynthesis can explain
for the first time the observed galactic chemical evolution of potassium and
other odd-Z elements \citep{Ritter:2018dma}.

An opportunity emerges in Canada now by connecting the advances in 3D stellar
hydro  with state-of the art explosion simulations of successful (e.g.,
\citealt{fernandez_2015}) and failed supernovae \citep{fernandez_2018},
exploring the potential of collapsars as key sites for the $r$-process
\citep{siegel_2019}, and diagnosing the progenitor, SN explosion mechanism and
engine with X-ray observations of SNRs (e.g., \citealt{Braun2019} -- see
Fig.~1). 

High-resolution X-ray spectroscopy is particularly needed to unveil the SN
explosion mechanism, as demonstrated in a brief Hitomi observation of the LMC
SNR N132D \citep{Hitomi2018}.  Furthermore, there is significant Canadian
expertise (both observational and theoretical) in studying the diversity of
compact stellar remnants, their evolution and their connection to their
supernovae (e.g., \citealt{Zhou2019,Hebbar2019,Rogers2016}).  The next leap in
probing nucleosynthesis yields in core-collapse supernovae and connecting the
SNR to its progenitor requires a combination of multi-wavelength,
high-resolution spectroscopy (particularly in the optical/UV and X-rays which
will be provided by JWST, CASTOR, XRISM and ATHENA), combined with improved
modelling in 3D (e.g., \citealt{Moumen2019}).

\subsection{Nuclear burning on neutron stars}

Thermonuclear reactions on neutron star surfaces, and in their upper crusts,
provides a rich regime of intriguing physics.  A recent, exciting example is
the theoretical discovery that cycles of electron capture and decay (Urca
reactions) should occur in neutron star crusts, as well as in the core, leading
to rapid cooling \citep{schatz_2014}. This  theoretical development contrasts
with recent observations which show, paradoxically, that neutron star crusts
require some source of variable extra nuclear heating that is not currently
understood, based on observations of the cooling of neutron stars after
extended periods of accretion (e.g. \citealt{deibel_2015}). Observational
progress requires high-sensitivity, high-spatial-resolution X-ray observations
of cooling neutron stars over periods of years to decades (see, e.g.,
\citealt{brown_2009}). 

Faster burning on neutron star surfaces generates X-ray "bursts", which show a
variety of unusual behaviours that we do not currently understand.  Burst
oscillations are variations, near (but often not exactly at) the neutron star
spin frequency, observed during parts of bursts. Although we have models
interpreting them, e.g. as burning fronts, some burst oscillations strongly
disagree with these models (e.g., \citealt{mahmoodifar_2019}).  Some bursts
show evidence of nuclear burning products in the best (though quite limited)
X-ray spectra, but the identity of the burning products cannot yet be
determined (e.g., \citealt{strohmayer_2019}). To make progress on understanding
nuclear burning in bursts, we need X-ray observatories with very large
effective areas (to catch many photons in the short bursts), and preferably
high spectral resolution (to resolve lines). NICER is making progress now, but
we need the higher effective area of ATHENA, eXTP, STROBE-X, and/or Colibr\`i
to solve some of these issues (\S\ref{s:recomm}).

\section{Experimental nuclear astrophysics}
\label{s:exp}

\subsection{Nuclear data for nucleosynthesis}
\label{Sec:TRIUMF-Astro}

TRIUMF ({\tt www.triumf.ca}) is Canada's particle accelerator centre in
Vancouver, BC. Several TRIUMF research scientists (B. Davids, I. Dillmann, R.
Kruecken, A. Kwiatkowski,  C. Ruiz) and Canadian University professors (e.g. A.
Chen, D. Muecher, R. Kanungo and G. Christian) conduct active nuclear
astrophysics programs at the TRIUMF-ISAC (radioactive beam) facility, as well
as at facilities abroad. The focus of the measurements are astrophysically
important reaction rates of stable and radioactive nuclei, as well as the
determination of  properties of exotic, short-lived nuclei like half-lives,
branching ratios, masses, and specific nuclear structure features far off
stability.

Almost all experimental setups at the ISAC facilities devote part of their
research time to help solving questions around the origin of the elements, with
focus on measuring important reaction rates for nova and X-ray burst
nucleosynthesis [e.g. $^{19}$Ne$(p,\gamma)$$^{20}$Na \citep{Wilkinson2017},
$^{38}$K$(p,\gamma)$$^{39}$Ca \citep{Lotay2016}, \citep{Christian2018} measured
with the DRAGON recoil separator], nuclear properties of short-lived isotopes
for the rapid neutron-capture ($r$) process [e.g. decay half-lives of
neutron-rich $^{128-130}$Cd \citep{Dunlop2016} measured with the GRIFFIN
spectrometer], and in the near future also the indirect determination of
neutron-capture cross sections of radioactive nuclei for the intermediate
neutron-capture ($i$) process [e.g. upcoming experiments with the EMMA recoil
mass spectrometer in 2020/21 to constrain neutron-rich
$^{137,139,142}$Cs$(n,\gamma)$ and $^{139}$Ba$(n,\gamma)$ reactions]. The
latter are crucial inputs in nucleosynthesis calculations from core-collapse
and neutron star merger simulations. 

The presently constructed new ``Advanced Rare IsotopE Laboratory"
(ARIEL\footnote{\tt https://fiveyearplan.triumf.ca/platforms/ariel/}) will
triple the radioactive beam capabilities at ISAC within the next decade.
Particularly the nuclear astrophysics program will greatly benefit from the
possibility to carry out longer beam times and to produce cleaner radioactive
isotope beams.

\subsection{Neutrino Astrophysics}

SNOLAB ({\tt www.snolab.ca}) is the Canadian underground science laboratory
specializing in neutrino and dark matter physics and is located in the Vale
Creighton Mine near Sudbury ON (TRIUMF scientists are also involved in the
domestic neutrino program at SNOLAB).

Canada has a dedicated neutrino detector for core-collapse supernovae (HALO,
\citealt{zuber_2015}), which is part of the global supernova early warning
system (SNEWS\footnote{\url{https://snews.bnl.gov}}).  A galactic core-collapse
supernova will result in a strong neutrino signal for many detectors, providing
information about the supernova engine which is not available through any other
means (gravitational waves probe other aspects of the explosion mechanism, and
photons can only be observed once the shock reaches the stellar surface).

Neutrinos from a Galactic supernova will also be observable with the new
generation SNO+ detector \citep{SNOplus2019a,SNOplus2019b} and the upcoming
``next Enriched Xenon Observatory" (nEXO). Canadian scientists are also
involved in other world-class international neutrino experiments, such as TK2
and the upcoming Hyper-Kamiokande in Japan. These detectors are also capable of
probing Solar neutrinos, which directly diagnose nuclear reactions in the core.

\vspace{-0.1in}
\section{Resources needed and recommendations}
\label{s:recomm}

\begin{enumerate}

\vspace{-0.05in}
\item {\bf Participation in observational surveys (transient and
non-transient)}. Sustaining Canadian leadership in this field requires
participation in international surveys and facilities including LSST, Gemini,
SKA, and next-generation X-ray instruments such as ATHENA. See also white
papers led by Ruan (E035), Spekkens (E042), C\^ot\'e (E045) and Hoffman (E048).  

\vspace{-0.05in}
\item {\bf More Niagara-scale computing power}. Taking the next step in
solving the theoretical challenges in the field will require more machines of
similar or higher caliber for large parallel jobs (see Box 3 below).

\vspace{-0.05in}
\item {\bf Collaborative grants.} As demonstrated in this WP, nuclear
astrophysics requires coordination and interplay between different sub-areas of
Astrophysics and also with Experimental Physics. The field would greatly
benefit from new funding programs for interdisciplinary work (e.g., funding
workshops, see Box 3 below).

\vspace{-0.05in}
\item {\bf Postdoctoral funding}.  The interdisciplinary approach needed to
tackle key questions in Nuclear Astrophysics needs to be reflected in the HQP
training. The JINA example (Box 3 below) shows that the most successful PDFs
have been exposed during their training to multiple aspects of nuclear
astrophysics research, and are therefore able to identify the best
opportunities and create new science outcomes. We recommend that the
collaborative nuclear astrophysics funding instrument proposed above also
includes a component for postdoctoral fellows that are specifically dedicated
to work in at least two complementary areas outlined in this WP, and under the
umbrella of the collaborative research grant.The CITA National Fellowship and
the NSERC Postdoctoral Fellowship, while important for attracting elite
researchers to Canada, are not sufficient to meet the needs of the field.  See
also white paper led by Ngo (E028).
    
\end{enumerate}




\vspace{-0.05in}
\begin{lrptextbox}[How does the proposed initiative result in fundamental or
transformational advances in our understanding of the Universe?] This research
program addresses key open questions in our understanding of the origin of the
elements. It combines detailed simulations of astrophysical sites in stars and
stellar explosions with nuclear physics experiments and theory, in order to
make predictions that are tested with large surveys and detailed spectroscopic
abundance observations. The research also relates to broader questions in
astronomy, such as the formation and evolution of the early universe and of
galaxies. Stellar objects are used as laboratories to study physics in extreme
regimes such as neutrinos in supernova explosions and/or strong gravity
environments where detectable gravitational waves are produced. 

\end{lrptextbox}

\begin{lrptextbox}[What are the main scientific risks and how will they be
mitigated?] The main risk involves \emph{not carrying out this research program
and/or falling behind international competition}. On the theoretical side,
ensuring that more computing power of Niagara scale or larger continues to be
provided by Compute Canada or some other national organization is key. On the
observational side, Canadian participation in large surveys like LSST or MSE,
funding CASTOR, ensuring the availability of target-of-opportunity capabilities
in existing and future Canadian telescopes, and having access to next
generation X-ray instruments like Athena will ensure that Canadian researchers
can carry out the groundbreaking observations needed to lead in the field.
Finally, having a separate postdoctoral funding program will allow many small
Canadian teams -- that rely mostly on students and/or external collaborations
-- to grow to the next level and compete on a level-playing field with European
or US groups.

\end{lrptextbox}

\begin{lrptextbox}[Is there the expectation of and capacity for Canadian scientific, technical or strategic leadership? ]
A substantial number of new hires at Canadian institutions in recent years have
led to nuclear astrophysics emerging as a significant Canadian strength. Our
expertise comprises most of the research areas needed to attack the most
exiting problems in the field. On the Astrophysics side (\S\ref{s:science1} and
\S\ref{s:science2}) this expertise resides in researchers at Universities
(including UVic, UManitoba, UToronto, McGill, UAlberta,
Perimeter/Waterloo/Guelph) and on the Nuclear and Neutrino Physics side
(\S\ref{s:exp}) in scientists at TRIUMF and SNOLAB. Canadian researchers have
been involved in the leadership of the US NSF Physics Frontiers Center JINA,
and in science collaborations such as NuGrid.  Several Canadian scientists
based overseas also have significant expertise in this field and often
collaborate with domestic researchers.

Canadian activities are presently held back from reaching their full potential
due to the lack of dedicated funding instruments that promote and enable the
required networking and coordination activities. In the US the NSF Physics
Frontier Center Joint Institute for Nuclear Astrophysics (JINA) has (with some
Canadian participation) been extremely successful in defining and driving the
field. As JINA is engaging into the fourth renewal campaign there is a great
opportunity to fund a complementary Canadian Virtual Centre that would be
closely connected to and leveraged off JINA. 

High performance computational resources are provided through Compute Canada
and regional consortia, as well as via international collaborations (such as
NERSC-DOE and NSF).  The Niagara supercomputer has been transformational for
enabling some of the key new results highlighted in this white paper, which
would not have been possible with previous-generation facilities. A new
community of computational astrophysics has emerged which is addressing
theoretical challenges on stellar evolution, gravitational wave Physics,
explosive events, and galaxy evolution. It is of paramount importance that
there is an appropriate succession path for Niagara, enhancing on the existing
capabilities according to technical opportunities. In order to maintain the
presently reached status, as an absolute minimum a next generation machine
needs to have three times the computing capability by 2023 and the capability
must be nine-fold by 2028. As a comparison, the 2019-launched NSF machine
Frontera (TACC) has 5.3 times as many nodes as Niagara, and each of them is
about 1.7 times more powerful.

\end{lrptextbox}

\newpage
\begin{lrptextbox}[Is there support from, involvement from, and coordination within the relevant Canadian community and more broadly?] 
The groups and communities represented in this WP are coordinating informally.
The community is eager to collaborate more effectively, but, as stated in the
previous box, what is missing is a collaborative funding instrument that
matches the needs of nuclear astrophysics. This would bring the  Canadian
collection of efforts at several institutions, including TRIUMF/SNOLAB and
Universities, and the associated international collaborators together, and
enhance the impact of Canadian participation in international projects, mostly
with the US, Europe, and Japan. 

\end{lrptextbox}

\begin{lrptextbox}[Will this program position Canadian astronomy for future opportunities and returns in 2020-2030 or beyond 2030?] 
Given existing expertise in the field, the investments that we propose will
position Canada as a leading player worldwide in the next decade: carrying out
the state-of-the-art numerical simulations needed to predict nucleosynthetic
yields from stellar to galactic scales, following up multi-messenger transients
at increasing detection rates, surveying the Galaxy and beyond to reconstruct
the chemical enrichment history of the Universe, and exploring the frontier
properties of nuclei and particles.  

\end{lrptextbox}

\begin{lrptextbox}[In what ways is the cost-benefit ratio, including existing investments and future operating costs, favourable?] 
Investment in more supercomputing capabilities of Niagara scale or larger will
not only benefit our field, but all other fields of Astrophysics (and other
sciences) in which significant parallel computing power (jobs of more than 1000
tasks) is required. Observational facilities and surveys are generally
multi-purpose, thus Canadian participation in any of them will provide benefits
beyond our subset of the community.  A new program for funding postdoctoral
researchers will help retain HQP in Canada that took significant investment to
train (instead of giving it away to our competitors or international
collaborators), and will also help attract international talent, bringing in
expertise that would otherwise not be found in Canada. 

\end{lrptextbox}

\begin{lrptextbox}[What are the main programmatic risks and how will they be mitigated?] 
Currently there is a disconnect between the nuclear/particle experiment side
and the astronomy side in Canada. We need resources and a platform to stimulate
and enable these inter-disciplinary collaborations.  New funding opportunities
for workshops will help gather these communities and provide a larger pool of
HQP to recruit from. Combining these types of events with a new program for
postdoctoral funding, we can transfer expertise from one community to the other
and maximize the benefits of existing investments by each community.
\end{lrptextbox}

\vspace{-0.1in}
\begin{lrptextbox}[Does the proposed initiative offer specific tangible benefits to Canadians, including but not limited to interdisciplinary research, industry opportunities, HQP training,
EDI,
outreach or education?] 

A key benefit of our recommendations is improvement in interdisciplinary
collaboration among the various sub-fields of astronomy involved and with the
nuclear / particle experimental community. This will strengthen HQP training in
a wide range of skills. In particular, our expertise in high-performance
computing, combined with the diversity of our group, can provide a fertile
ground for increasing the participation of under-represented groups in
computing-intensive areas such as Data Science.  

\end{lrptextbox}

\bibliography{nuclear_astro} 

\begin{thebibliography}{}
\expandafter\ifx\csname natexlab\endcsname\relax\def\natexlab#1{#1}\fi

\bibitem[{{Abbott} {et~al.}(2017)}]{gw170817_mm}
{Abbott}, B.~P., {et~al.} 2017, \apjl, 848, L12

\bibitem[{{Abbott} {et~al.}(2018)}]{ligo_obs_plans}
---. 2018, Living Reviews in Relativity, 21, 3

\bibitem[{Andrassy {et~al.}(2018)Andrassy, Herwig, Woodward, \&
  Ritter}]{Andrassy:2018wy}
Andrassy, R., Herwig, F., Woodward, P., \& Ritter, C. 2018, MNRAS, submitted,
  arXiv:1808.04014

\bibitem[{{Armillotta} {et~al.}(2018)}]{armillota18}
{Armillotta}, L., {et~al.} 2018, \mnras, 481, 5000

\bibitem[{{Asplund} {et~al.}(2009){Asplund}, {Grevesse}, {Sauval}, \&
  {Scott}}]{asplund09}
{Asplund}, M., {Grevesse}, N., {Sauval}, A.~J., \& {Scott}, P. 2009, \araa, 47,
  481

\bibitem[{{Blair} \& {Raymond}(2017)}]{BR17}
{Blair}, W.~P., \& {Raymond}, J.~C. 2017, {Ultraviolet and Optical Insights
  into Supernova Remnant Shocks}, 2087

\bibitem[{Bowman {et~al.}(2019)Bowman, Burssens, Pedersen, Johnston, Aerts,
  Buysschaert, Michielsen, Tkachenko, Rogers, Edelmann, Ratnasingam,
  Sim{\'o}n-D{\'\i}az, Castro, Moravveji, Pope, White, \&
  De~Cat}]{Bowman:2019ka}
Bowman, D.~M., Burssens, S., Pedersen, M.~G., {et~al.} 2019, Nature Astronomy
  2018, 528, 1

\bibitem[{{Braun} {et~al.}(2019){Braun}, {Safi-Harb}, \& {Fryer}}]{Braun2019}
{Braun}, C., {Safi-Harb}, S., \& {Fryer}, C.~L. 2019, \mnras, 2103

\bibitem[{{Brown} \& {Cumming}(2009)}]{brown_2009}
{Brown}, E.~F., \& {Cumming}, A. 2009, \apj, 698, 1020

\bibitem[{{Chen} {et~al.}(2014){Chen}, {Woods}, {Yungelson}, {Gilfanov}, \&
  {Han}}]{Chen14}
{Chen}, H.-L., {Woods}, T.~E., {Yungelson}, L.~R., {Gilfanov}, M., \& {Han}, Z.
  2014, \mnras, 445, 1912

\bibitem[{Christian {et~al.}(2018)Christian, Lotay, Ruiz, Akers, Burke,
  Catford, Chen, Connolly, Davids, Fallis, Hager, Hutcheon, Mahl, Rojas, \&
  Sun}]{Christian2018}
Christian, G., Lotay, G., Ruiz, C., {et~al.} 2018, Phys. Rev. C, 97, 025802

\bibitem[{{C{\^o}t{\'e}} {et~al.}(2018{\natexlab{a}}){C{\^o}t{\'e}},
  {Denissenkov}, {Herwig}, {Ruiter}, {Ritter}, {Pignatari}, \&
  {Belczynski}}]{cote18_iprocess}
{C{\^o}t{\'e}}, B., {Denissenkov}, P., {Herwig}, F., {et~al.}
  2018{\natexlab{a}}, \apj, 854, 105

\bibitem[{{C{\^o}t{\'e}} {et~al.}(2018{\natexlab{b}}){C{\^o}t{\'e}}, {Fryer},
  {Belczynski}, {Korobkin}, {Chru{\'s}li{\'n}ska}, {Vassh}, {Mumpower},
  {Lippuner}, {Sprouse}, {Surman}, \& {Wollaeger}}]{cote18_ligo}
{C{\^o}t{\'e}}, B., {Fryer}, C.~L., {Belczynski}, K., {et~al.}
  2018{\natexlab{b}}, \apj, 855, 99

\bibitem[{{Dave} {et~al.}(2017){Dave}, {Kashyap}, {Fisher}, {Timmes},
  {Townsley}, \& {Byrohl}}]{Dave17}
{Dave}, P., {Kashyap}, R., {Fisher}, R., {et~al.} 2017, \apj, 841, 58

\bibitem[{{Deibel} {et~al.}(2015){Deibel}, {Cumming}, {Brown}, \&
  {Page}}]{deibel_2015}
{Deibel}, A., {Cumming}, A., {Brown}, E.~F., \& {Page}, D. 2015, \apjl, 809,
  L31

\bibitem[{{Denissenkov} {et~al.}(2017){Denissenkov}, {Herwig}, {Battino},
  {Ritter}, {Pignatari}, {Jones}, \& {Paxton}}]{Denissenkov17}
{Denissenkov}, P.~A., {Herwig}, F., {Battino}, U., {et~al.} 2017, \apjl, 834,
  L10

\bibitem[{Denissenkov {et~al.}(2019)Denissenkov, Herwig, Woodward, Andrassy,
  Pignatari, \& Jones}]{2019MNRAS.488.4258D}
Denissenkov, P.~A., Herwig, F., Woodward, P., {et~al.} 2019, MNRAS, 488, 4258

\bibitem[{{Drout} {et~al.}(2017){Drout}, {Piro}, {Shappee}, {Kilpatrick},
  {Simon}, {Contreras}, {Coulter}, {Foley}, {Siebert}, {Morrell}, {Boutsia},
  {Di Mille}, {Holoien}, {Kasen}, {Kollmeier}, {Madore}, {Monson},
  {Murguia-Berthier}, {Pan}, {Prochaska}, {Ramirez-Ruiz}, {Rest}, {Adams},
  {Alatalo}, {Ba{\~n}ados}, {Baughman}, {Beers}, {Bernstein}, {Bitsakis},
  {Campillay}, {Hansen}, {Higgs}, {Ji}, {Maravelias}, {Marshall}, {Moni Bidin},
  {Prieto}, {Rasmussen}, {Rojas-Bravo}, {Strom}, {Ulloa},
  {Vargas-Gonz{\'a}lez}, {Wan}, \& {Whitten}}]{drout_2017}
{Drout}, M.~R., {Piro}, A.~L., {Shappee}, B.~J., {et~al.} 2017, Science, 358,
  1570

\bibitem[{Dunlop {et~al.}(2016)Dunlop, Bildstein, Dillmann, Jungclaus,
  Svensson, Andreoiu, Ball, Bernier, Bidaman, Boubel, Burbadge,
  Caballero-Folch, Dunlop, Evitts, Garcia, Garnsworthy, Garrett, Hackman,
  Hallam, Henderson, Ilyushkin, Kisliuk, Kruecken, Lassen, Li, MacConnachie,
  MacLean, McGee, Moukaddam, Olaizola, E., Park, Paetkau, Petrache, Pore,
  Radich, Ruotsalainen, Smallcombe, Smith, Tabor, Teigelh\"{o}fer, Turko, \&
  Zidar}]{Dunlop2016}
Dunlop, R., Bildstein, V., Dillmann, I., {et~al.} 2016, Phys. Rev. C, 93,
  062801(R)

\bibitem[{{Fern{\'a}ndez}(2015)}]{fernandez_2015}
{Fern{\'a}ndez}, R. 2015, \mnras, 452, 2071

\bibitem[{{Fern{\'a}ndez} {et~al.}(2017){Fern{\'a}ndez}, {Foucart}, {Kasen},
  {Lippuner}, {Desai}, \& {Roberts}}]{fernandez_2017}
{Fern{\'a}ndez}, R., {Foucart}, F., {Kasen}, D., {et~al.} 2017, Classical and
  Quantum Gravity, 34, 154001

\bibitem[{{Fern{\'a}ndez} {et~al.}(2018){Fern{\'a}ndez}, {Quataert},
  {Kashiyama}, \& {Coughlin}}]{fernandez_2018}
{Fern{\'a}ndez}, R., {Quataert}, E., {Kashiyama}, K., \& {Coughlin}, E.~R.
  2018, \mnras, 476, 2366

\bibitem[{{Frebel} \& {Norris}(2015)}]{frebel15_review}
{Frebel}, A., \& {Norris}, J.~E. 2015, \araa, 53, 631

\bibitem[{{Graur} \& {Woods}(2019)}]{GW19}
{Graur}, O., \& {Woods}, T.~E. 2019, \mnras, 484, L79

\bibitem[{{Haggard} {et~al.}(2017){Haggard}, {Nynka}, {Ruan}, {Kalogera},
  {Cenko}, {Evans}, \& {Kennea}}]{haggard_2017}
{Haggard}, D., {Nynka}, M., {Ruan}, J.~J., {et~al.} 2017, \apjl, 848, L25

\bibitem[{{Hebbar} {et~al.}(2019){Hebbar}, {Heinke}, \& {Ho}}]{Hebbar2019}
{Hebbar}, P.~R., {Heinke}, C.~O., \& {Ho}, W. C.~G. 2019, \mnras, 2213

\bibitem[{Herwig {et~al.}(2014)Herwig, Woodward, Lin, Knox, \&
  Fryer}]{Herwig:2014cx}
Herwig, F., Woodward, P.~R., Lin, P.-H., Knox, M., \& Fryer, C. 2014, ApJ, 792,
  L3

\bibitem[{{Hitomi Collaboration} {et~al.}(2017){Hitomi Collaboration},
  {Aharonian}, {Akamatsu}, {Akimoto}, {Allen}, {Angelini}, {Audard}, {Awaki},
  {Axelsson}, {Bamba}, {Bautz}, {Blandford}, {Brenneman}, {Brown}, {Bulbul},
  {Cackett}, {Chernyakova}, {Chiao}, {Coppi}, {Costantini}, {de Plaa}, {den
  Herder}, {Done}, {Dotani}, {Ebisawa}, {Eckart}, {Enoto}, {Ezoe}, {Fabian},
  {Ferrigno}, {Foster}, {Fujimoto}, {Fukazawa}, {Furuzawa}, {Galeazzi},
  {Gallo}, {Gandhi}, {Giustini}, {Goldwurm}, {Gu}, {Guainazzi}, {Haba},
  {Hagino}, {Hamaguchi}, {Harrus}, {Hatsukade}, {Hayashi}, {Hayashi},
  {Hayashida}, {Hiraga}, {Hornschemeier}, {Hoshino}, {Hughes}, {Ichinohe},
  {Iizuka}, {Inoue}, {Inoue}, {Ishida}, {Ishikawa}, {Ishisaki}, {Iwai},
  {Kaastra}, {Kallman}, {Kamae}, {Kataoka}, {Katsuda}, {Kawai}, {Kelley},
  {Kilbourne}, {Kitaguchi}, {Kitamoto}, {Kitayama}, {Kohmura}, {Kokubun},
  {Koyama}, {Koyama}, {Kretschmar}, {Krimm}, {Kubota}, {Kunieda}, {Laurent},
  {Lee}, {Leutenegger}, {Limousine}, {Loewenstein}, {Long}, {Lumb}, {Madejski},
  {Maeda}, {Maier}, {Makishima}, {Markevitch}, {Matsumoto}, {Matsushita},
  {McCammon}, {McNamara}, {Mehdipour}, {Miller}, {Miller}, {Mineshige},
  {Mitsuda}, {Mitsuishi}, {Miyazawa}, {Mizuno}, {Mori}, {Mori}, {Mukai},
  {Murakami}, {Mushotzky}, {Nakagawa}, {Nakajima}, {Nakamori}, {Nakashima},
  {Nakazawa}, {Nobukawa}, {Nobukawa}, {Noda}, {Odaka}, {Ohashi}, {Ohno},
  {Okajima}, {Ota}, {Ozaki}, {Paerels}, {Paltani}, {Petre}, {Pinto}, {Porter},
  {Pottschmidt}, {Reynolds}, {Safi-Harb}, {Saito}, {Sakai}, {Sasaki}, {Sato},
  {Sato}, {Sato}, {Sawada}, {Schartel}, {Serlemitsos}, {Seta}, {Shidatsu},
  {Simionescu}, {Smith}, {Soong}, {Stawarz}, {Sugawara}, {Sugita},
  {Szymkowiak}, {Tajima}, {Takahashi}, {Takahashi}, {Takeda}, {Takei},
  {Tamagawa}, {Tamura}, {Tanaka}, {Tanaka}, {Tanaka}, {Tashiro}, {Tawara},
  {Terada}, {Terashima}, {Tombesi}, {Tomida}, {Tsuboi}, {Tsujimoto}, {Tsunemi},
  {Go Tsuru}, {Uchida}, {Uchiyama}, {Uchiyama}, {Ueda}, {Ueda}, {Uno}, {Urry},
  {Ursino}, {de Vries}, {Watanabe}, {Werner}, {Wik}, {Wilkins}, {Williams},
  {Yamada}, {Yamaguchi}, {Yamaoka}, {Yamasaki}, {Yamauchi}, {Yamauchi},
  {Yaqoob}, {Yatsu}, {Yonetoku}, {Zhuravleva}, \& {Zoghbi}}]{Hitomi2017}
{Hitomi Collaboration}, {Aharonian}, F., {Akamatsu}, H., {et~al.} 2017, \nat,
  551, 478

\bibitem[{{Hitomi Collaboration} {et~al.}(2018){Hitomi Collaboration},
  {Aharonian}, {Akamatsu}, {Akimoto}, {Allen}, {Angelini}, {Audard}, {Awaki},
  {Axelsson}, {Bamba}, {Bautz}, {Blandford}, {Brenneman}, {Brown}, {Bulbul},
  {Cackett}, {Chernyakova}, {Chiao}, {Coppi}, {Costantini}, {de Plaa}, {de
  Vries}, {den Herder}, {Done}, {Dotani}, {Ebisawa}, {Eckart}, {Enoto}, {Ezoe},
  {Fabian}, {Ferrigno}, {Foster}, {Fujimoto}, {Fukazawa}, {Furuzawa},
  {Galeazzi}, {Gallo}, {Gandhi}, {Giustini}, {Goldwurm}, {Gu}, {Guainazzi},
  {Haba}, {Hagino}, {Hamaguchi}, {Harrus}, {Hatsukade}, {Hayashi}, {Hayashi},
  {Hayashida}, {Hiraga}, {Hornschemeier}, {Hoshino}, {Hughes}, {Ichinohe},
  {Iizuka}, {Inoue}, {Inoue}, {Ishida}, {Ishikawa}, {Ishisaki}, {Iwai},
  {Kaastra}, {Kallman}, {Kamae}, {Kataoka}, {Katsuda}, {Kawai}, {Kelley},
  {Kilbourne}, {Kitaguchi}, {Kitamoto}, {Kitayama}, {Kohmura}, {Kokubun},
  {Koyama}, {Koyama}, {Kretschmar}, {Krimm}, {Kubota}, {Kunieda}, {Laurent},
  {Lee}, {Leutenegger}, {Limousin}, {Loewenstein}, {Long}, {Lumb}, {Madejski},
  {Maeda}, {Maier}, {Makishima}, {Markevitch}, {Matsumoto}, {Matsushita},
  {McCammon}, {McNamara}, {Mehdipour}, {Miller}, {Miller}, {Mineshige},
  {Mitsuda}, {Mitsuishi}, {Miyazawa}, {Mizuno}, {Mori}, {Mori}, {Mukai},
  {Murakami}, {Mushotzky}, {Nakagawa}, {Nakajima}, {Nakamori}, {Nakashima},
  {Nakazawa}, {Nobukawa}, {Nobukawa}, {Noda}, {Odaka}, {Ohashi}, {Ohno},
  {Okajima}, {Ota}, {Ozaki}, {Paerels}, {Paltani}, {Petre}, {Pinto}, {Porter},
  {Pottschmidt}, {Reynolds}, {Safi-Harb}, {Saito}, {Sakai}, {Sasaki}, {Sato},
  {Sato}, {Sato}, {Sato}, {Sawada}, {Schartel}, {Serlemtsos}, {Seta},
  {Shidatsu}, {Simionescu}, {Smith}, {Soong}, {Stawarz}, {Sugawara}, {Sugita},
  {Szymkowiak}, {Tajima}, {Takahashi}, {Takahashi}, {Takeda}, {Takei},
  {Tamagawa}, {Tamura}, {Tanaka}, {Tanaka}, {Tanaka}, {Tashiro}, {Tawara},
  {Terada}, {Terashima}, {Tombesi}, {Tomida}, {Tsuboi}, {Tsujimoto}, {Tsunemi},
  {Tsuru}, {Uchida}, {Uchiyama}, {Uchiyama}, {Ueda}, {Ueda}, {Uno}, {Urry},
  {Ursino}, {Watanabe}, {Werner}, {Wilkins}, {Williams}, {Yamada}, {Yamaguchi},
  {Yamaoka}, {Yamasaki}, {Yamauchi}, {Yamauchi}, {Yaqoob}, {Yatsu}, {Yonetoku},
  {Zhuravleva}, \& {Zoghbi}}]{Hitomi2018}
---. 2018, \pasj, 70, 16

\bibitem[{{Hopkins} {et~al.}(2018){Hopkins}, {Wetzel}, {Kere{\v s}},
  {Faucher-Gigu{\`e}re}, {Quataert}, {Boylan-Kolchin}, {Murray}, {Hayward},
  {Garrison-Kimmel}, {Hummels}, {Feldmann}, {Torrey}, {Ma},
  {Angl{\'e}s-Alc{\'a}zar}, {Su}, {Orr}, {Schmitz}, {Escala}, {Sanderson},
  {Grudi{\'c}}, {Hafen}, {Kim}, {Fitts}, {Bullock}, {Wheeler}, {Chan},
  {Elbert}, \& {Narayanan}}]{hopkins18}
{Hopkins}, P.~F., {Wetzel}, A., {Kere{\v s}}, D., {et~al.} 2018, \mnras, 480,
  800

\bibitem[{{Ivanova} {et~al.}(2013){Ivanova}, {Justham}, {Chen}, {De Marco},
  {Fryer}, {Gaburov}, {Ge}, {Glebbeek}, {Han}, {Li}, {Lu}, {Marsh},
  {Podsiadlowski}, {Potter}, {Soker}, {Taam}, {Tauris}, {van den Heuvel}, \&
  {Webbink}}]{Ivanova13}
{Ivanova}, N., {Justham}, S., {Chen}, X., {et~al.} 2013, \aapr, 21, 59

\bibitem[{{Janka} {et~al.}(2016){Janka}, {Melson}, \& {Summa}}]{janka_2016}
{Janka}, H.-T., {Melson}, T., \& {Summa}, A. 2016, Annual Review of Nuclear and
  Particle Science, 66, 341

\bibitem[{Jones {et~al.}(2017)Jones, Andr{\'a}ssy, Sandalski, Davis, Woodward,
  \& Herwig}]{Jones:2017kc}
Jones, S., Andr{\'a}ssy, R., Sandalski, S., {et~al.} 2017, MNRAS, 465, 2991

\bibitem[{Kreckel {et~al.}(2019)Kreckel, Ho, Blanc, {et~al.}}]{kreckel_2019}
Kreckel, K., Ho, I.~T., Blanc, G.~A., {et~al.} 2019, ApJ, in press,
  arXiv:1910.07190

\bibitem[{{Lattimer} \& {Schramm}(1974)}]{lattimer_1974}
{Lattimer}, J.~M., \& {Schramm}, D.~N. 1974, \apjl, 192, L145

\bibitem[{{Lehner} {et~al.}(2016){Lehner}, {Liebling}, {Palenzuela},
  {Caballero}, {O'Connor}, {Anderson}, \& {Neilsen}}]{lehner_2016}
{Lehner}, L., {Liebling}, S.~L., {Palenzuela}, C., {et~al.} 2016, Classical and
  Quantum Gravity, 33, 184002

\bibitem[{Lotay {et~al.}(2016)Lotay, Christian, Ruiz, Akers, Burke, Catford,
  Chen, Connolly, Davids, Fallis, Hager, Hutcheon, Mahl, Rojas, \&
  Sun}]{Lotay2016}
Lotay, G., Christian, G., Ruiz, C., {et~al.} 2016, Phys. Rev. Lett., 116,
  132701

\bibitem[{{Mackereth} {et~al.}(2019){Mackereth}, {Schiavon}, {Pfeffer},
  {Hayes}, {Bovy}, {Anguiano}, {Allende Prieto}, {Hasselquist}, {Holtzman},
  {Johnson}, {Majewski}, {O'Connell}, {Shetrone}, {Tissera}, \&
  {Fern{\'a}ndez-Trincado}}]{mackereth19}
{Mackereth}, J.~T., {Schiavon}, R.~P., {Pfeffer}, J., {et~al.} 2019, \mnras,
  482, 3426

\bibitem[{{Mahmoodifar} {et~al.}(2019){Mahmoodifar}, {Strohmayer}, {Bult},
  {Altamirano}, {Arzoumanian}, {Chakrabarty}, {Gendreau}, {Guillot}, {Homan},
  {Jaisawal}, {Keek}, \& {Wolff}}]{mahmoodifar_2019}
{Mahmoodifar}, S., {Strohmayer}, T.~E., {Bult}, P., {et~al.} 2019, \apj, 878,
  145

\bibitem[{{Maoz} {et~al.}(2014){Maoz}, {Mannucci}, \& {Nelemans}}]{MMN14}
{Maoz}, D., {Mannucci}, F., \& {Nelemans}, G. 2014, \araa, 52, 107

\bibitem[{{Moumen} {et~al.}(2019){Moumen}, {Robert}, {Devost}, {Martin},
  {Rousseau-Nepton}, {Drissen}, \& {Martin}}]{Moumen2019}
{Moumen}, I., {Robert}, C., {Devost}, D., {et~al.} 2019, \mnras, 488, 803

\bibitem[{{MSE Science Team} {et~al.}(2019){MSE Science Team}, {Babusiaux},
  {Bergemann}, {Burgasser}, {Ellison}, {Haggard}, {Huber}, {Kaplinghat}, {Li},
  {Marshall}, \& et~al.}]{mse19}
{MSE Science Team}, {Babusiaux}, C., {Bergemann}, M., {et~al.} 2019, preprint,
  arXiv:1904.04907

\bibitem[{{Nomoto} {et~al.}(2013){Nomoto}, {Kobayashi}, \&
  {Tominaga}}]{nomoto13}
{Nomoto}, K., {Kobayashi}, C., \& {Tominaga}, N. 2013, \araa, 51, 457

\bibitem[{{Pazder} {et~al.}(2016){Pazder}, {Burley}, {Ireland}, {Robertson},
  {Sheinis}, \& {Zhelem}}]{GHOST2016}
{Pazder}, J., {Burley}, G., {Ireland}, M.~J., {et~al.} 2016, in Society of
  Photo-Optical Instrumentation Engineers (SPIE) Conference Series, Vol. 9908,
  \procspie, 99087F

\bibitem[{{Rennehan} {et~al.}(2019){Rennehan}, {Babul}, {Hopkins}, {Dav{\'e}},
  \& {Moa}}]{rennehan19}
{Rennehan}, D., {Babul}, A., {Hopkins}, P.~F., {Dav{\'e}}, R., \& {Moa}, B.
  2019, \mnras, 483, 3810

\bibitem[{Ritter {et~al.}(2018{\natexlab{a}})Ritter, Andr{\'a}ssy,
  C{\^o}t{\'e}, Herwig, Woodward, Pignatari, \& Jones}]{Ritter:2018dma}
Ritter, C., Andr{\'a}ssy, R., C{\^o}t{\'e}, B., {et~al.} 2018{\natexlab{a}},
  MNRAS, 474, L1

\bibitem[{Ritter {et~al.}(2018{\natexlab{b}})Ritter, Herwig, Jones, Pignatari,
  Fryer, \& Hirschi}]{Ritter:2018kb}
Ritter, C., Herwig, F., Jones, S., {et~al.} 2018{\natexlab{b}}, MNRAS, 480, 538

\bibitem[{{Rogers} \& {Safi-Harb}(2016)}]{Rogers2016}
{Rogers}, A., \& {Safi-Harb}, S. 2016, \mnras, 457, 1180

\bibitem[{{Rousseau-Nepton} {et~al.}(2019){Rousseau-Nepton}, {Martin},
  {Robert}, {et~al.}}]{signals}
{Rousseau-Nepton}, L., {Martin}, R.~P., {Robert}, C., {et~al.} 2019, MNRAS,
  submitted, arXiv:1908.09017

\bibitem[{{Schatz} {et~al.}(2014)}]{schatz_2014}
{Schatz}, H., {et~al.} 2014, \nat, 505, 62

\bibitem[{{Searle}(1971)}]{searle71}
{Searle}, L. 1971, \apj, 168, 327

\bibitem[{{Seitenzahl} \& {Townsley}(2017)}]{ST17}
{Seitenzahl}, I.~R., \& {Townsley}, D.~M. 2017, {Nucleosynthesis in
  Thermonuclear Supernovae}, 1955

\bibitem[{{Siegel} {et~al.}(2019){Siegel}, {Barnes}, \&
  {Metzger}}]{siegel_2019}
{Siegel}, D.~M., {Barnes}, J., \& {Metzger}, B.~D. 2019, \nat, 569, 241

\bibitem[{{Siegel} \& {Metzger}(2017)}]{siegel_2017}
{Siegel}, D.~M., \& {Metzger}, B.~D. 2017, \prl, 119, 231102

\bibitem[{{Starkenburg} {et~al.}(2017){Starkenburg}, {Oman}, {Navarro},
  {Crain}, {Fattahi}, {Frenk}, {Sawala}, \& {Schaye}}]{starkenburg17}
{Starkenburg}, E., {Oman}, K.~A., {Navarro}, J.~F., {et~al.} 2017, \mnras, 465,
  2212

\bibitem[{{Starkenburg} {et~al.}(2018){Starkenburg}, {Aguado}, {Bonifacio},
  {Caffau}, {Jablonka}, {Lardo}, {Martin}, {S{\'a}nchez-Janssen}, {Sestito},
  {Venn}, {Youakim}, {Allende Prieto}, {Arentsen}, {Gentile}, {Gonz{\'a}lez
  Hern{\'a}ndez}, {Kielty}, {Koppelman}, {Longeard}, {Tolstoy}, {Carlberg},
  {C{\^o}t{\'e}}, {Fouesneau}, {Hill}, {McConnachie}, \&
  {Navarro}}]{pristineIV}
{Starkenburg}, E., {Aguado}, D.~S., {Bonifacio}, P., {et~al.} 2018, \mnras,
  481, 3838

\bibitem[{{Strohmayer} {et~al.}(2019)}]{strohmayer_2019}
{Strohmayer}, T.~E., {et~al.} 2019, \apjl, 878, L27

\bibitem[{{The SNO+ collaboration}(2019{\natexlab{a}})}]{SNOplus2019b}
{The SNO+ collaboration}. 2019{\natexlab{a}}, Phys. Rev. D, 99, 012012

\bibitem[{{The SNO+ collaboration}(2019{\natexlab{b}})}]{SNOplus2019a}
---. 2019{\natexlab{b}}, Phys. Rev. D, 99, 032008

\bibitem[{{Thorp} {et~al.}(2019)}]{thorp19}
{Thorp}, M.~D., {et~al.} 2019, \mnras, 482, L55

\bibitem[{{Venn} {et~al.}(2019){Venn}, {Kielty}, {Sestito},
  {et~al.}}]{venn2019}
{Venn}, K., {Kielty}, C., {Sestito}, F., {et~al.} 2019, MNRAS, submitted,
  arXiv:1910.06340

\bibitem[{{Venn} {et~al.}(2004){Venn}, {Irwin}, {Shetrone}, {Tout}, {Hill}, \&
  {Tolstoy}}]{venn04}
{Venn}, K.~A., {Irwin}, M., {Shetrone}, M.~D., {et~al.} 2004, \aj, 128, 1177

\bibitem[{Wilkinson {et~al.}(2017)Wilkinson, Lotay, Lennarz, Ruiz, Christian,
  Akers, Catford, Chen, Connolly, Davids, Hutcheon, Jedrejcic, Laird, Martin,
  McNeice, Riley, \& Williams}]{Wilkinson2017}
Wilkinson, R., Lotay, G., Lennarz, A., {et~al.} 2017, Phys. Rev. Lett., 119,
  242701

\bibitem[{{Wilson} \& {Rood}(1994)}]{wilson94}
{Wilson}, T.~L., \& {Rood}, R. 1994, \araa, 32, 191

\bibitem[{{Woods} {et~al.}(2017){Woods}, {Ghavamian}, {Badenes}, \&
  {Gilfanov}}]{Woods17}
{Woods}, T.~E., {Ghavamian}, P., {Badenes}, C., \& {Gilfanov}, M. 2017, Nature
  Astronomy, 1, 800

\bibitem[{{Zhou} {et~al.}(2019){Zhou}, {Vink}, {Safi-Harb}, \&
  {Miceli}}]{Zhou2019}
{Zhou}, P., {Vink}, J., {Safi-Harb}, S., \& {Miceli}, M. 2019, \aap, 629, A51

\bibitem[{{Zhu} {et~al.}(2015){Zhu}, {Pakmor}, {van Kerkwijk}, \&
  {Chang}}]{Zhu15}
{Zhu}, C., {Pakmor}, R., {van Kerkwijk}, M.~H., \& {Chang}, P. 2015, \apjl,
  806, L1

\bibitem[{{Zuber}(2015)}]{zuber_2015}
{Zuber}, K. 2015, Nuclear and Particle Physics Proceedings, 265, 233

\end{thebibliography}

\end{document}